\documentclass[twocolumn,preprintnumbers]{revtex4}
\usepackage{graphicx}
\usepackage{amsmath}
\usepackage{amssymb}
\usepackage{amsthm}
\usepackage{amsfonts}
\usepackage{enumerate}
\usepackage{centernot}
\usepackage{tikz}

\input{tcilatex}
\begin{document}

\title{First Principles Study of the Giant Magnetic Anisotropy Energy in
Bulk Na$_{4}$IrO$_{4}$}
\author{Di Wang$^{1}$, Feng Tang$^{1}$, Yongping Du$^{2,1}$, and Xiangang Wan%
$^{1}$}
\thanks{Corresponding author: xgwan@nju.edu.cn}
\affiliation{$^{1}$National Laboratory of Solid State Microstructures and Department of
Physics, Nanjing University, Nanjing 210093, China\\
$^{2}$Department of Applied Physics, Nanjing University of Science and
Technology, Nanjing 210094, China}

\begin{abstract}
In $5d$ transition metal oxides, novel properties arise from the interplay
of electron correlations and spin--orbit interactions. Na$_{4}$IrO$_{4}$,
where $5d$ transition-metal Ir atom occupies the center of the square-planar
coordination environment, is synthesized. Based on density functional
theory, we calculate its electronic and magnetic properties. Our numerical
results show that the Ir-$5d$ bands are quite narrow, and the bands around
the Fermi level are mainly contributed by $d_{xy},d_{yz}$ and $d_{zx}$
orbitals. The magnetic easy-axis is perpendicular to the IrO$_{4}$ plane,
and the magnetic anisotropy energy (MAE) of Na$_{4}$IrO$_{4}$ is found to be
very giant. We estimate the magnetic parameters by mapping the calculated
total energy for different spin configurations onto a spin model. The next
nearest neighbor exchange interaction $J_{2}$ is much larger than other
intersite exchange interactions and results in the magnetic ground state
configuration. Our study clearly demonstrates that the huge MAE comes from
the single-ion anisotropy rather than the anisotropic interatomic spin
exchange. This compound has a large spin gap but very narrow spin-wave
dispersion, due to the large single-ion anisotropy and relatively small
exchange couplings. Noticing this remarkable magnetic feature originated
from its highly isolated IrO$_{4}$ moiety, we also explore the possiblity to
further enhance the MAE.
\end{abstract}

\date{\today }
\maketitle

\section{Introduction}

It is well known that the Coulomb interaction is of substantial importance
in $3d$ electron systems, while the spin-orbit coupling (SOC) in these
compounds is quite small \cite{3d-1,3d-2}. However, the SOC and electronic
correlation in $5d$ electrons have comparable magnitudes. The delicate
interplay between electronic interactions, strong SOC, and crystal field
splitting can result in strongly competing ground states in these materials 
\cite{Rev-5d-1,Rev-5d-2,Rev-5d-3}. Thus recently, $5d$ transition metal
(especially Ir or Os) oxides have attracted intensive interest and a great
number of exotic phenomena have been observed experimentally or proposed
theoretically, e.g. $J_{eff}$=1/2 Mott state \cite{J-1/2 1,J-1/2 2,J-1/2 3},
topological insulator \cite{TI-5d-1,TI-5d-2,TI-5d-3}, Kitaev model \cite%
{Kitav}, Weyl Semimetal \cite{WSM}, high $T_{c}$ superconductivity \cite%
{HTC-214}, Axion insulator \cite{Axion-Insulator}, quantum spin liquid \cite%
{QSL-1,QSL-2}, Slater insulator \cite{Slater-1,Slater-2,Slater-3},
ferroelectric metal \cite{FE-Metal-1,FE-Metal-2}, etc.

In all the aforementioned systems, the $5d$ ions lie in the octahedral
environment of the O ions. In addition to this common coordination geometry,
Na$_{4}$IrO$_{4}$, where Ir atom occupies the center of the square-planar
coordination environment, has also been synthesized \cite{str}. By using
density function theory (DFT) calculations, Kanungo \cite{414} \textit{et al.%
} reveal that the relative weak Coulomb repulsion of Ir ions plays a key
role in the stabilization of the ideal square-planar geometry of the IrO$%
_{4} $ moiety in Na$_{4}$IrO$_{4}$. Located at the center of an ideally
square-planar IrO$_{4}$ oxoanion, the $5d$ electrons of Ir ions in Na$_{4}$%
IrO$_{4}$\ do not display the $J_{eff}$=1/2 configuration \cite{414}.
Moreover, the common $5d$ transition metal oxides own face (or edge,
corner)-sharing structure of oxygen octahedrons, while the square-planar IrO$%
_{4}$ oxoanion in Na$_{4}$IrO$_{4}$\ is quite isolated. Therefore, exploring
the possible exotic properties of Na$_{4}$IrO$_{4}$ is an interesting study.

In this work, based on first principle calculations, we systematically study
the electronic and magnetic properties of Na$_{4}$IrO$_{4}$. Our numerical
results show that the Ir-$5d$ bands are quite narrow, and the bands around
the Fermi level are mainly contributed by $d_{xy}$, $d_{yz}$ and $d_{zx}$
orbitals. Due to the isolated IrO$_{4}$\ moiety, the magnetic moments are
quite localized. We calculate several magnetic structures and find that the
antiferromagnetic-1 (AFM-1) state as shown in Fig. \ref{magneticstructure}
is the ground state configuration. The interatomic exchange interactions are
estimated and the nearest-neighbor $J_{2}$\ (shown in Fig. \ref{structure})
dominates over the others. We find that there is a huge magnetic anisotropy
energy (MAE) due to the special square-planar coordination environment and
the long distance between IrO$_{4}$\ moieties. We find the anisotropy of
interatomic spin exchange couplings is relatively small, and the huge MAE
comes from the single-ion anisotropy. We suggest that substituting Ir by Re
atom can further enhance the MAE significantly.

\section{Method and Crystal structure}

The electronic band structure and density of states calculations have been
carried out by using the full potential linearized augmented plane wave
method as implemented in Wien2k package \cite{wien2k}. Local spin density
approximation (LSDA) is widely used for various $4d$ and $5d$ TMOs \cite%
{J-1/2 1,J-1/2 2,WSM,Axion-Insulator,LSDA-4d}, and we therefore adopt it as
the exchange-correlation potential. A 9$\times $9$\times $15 k-point mesh is
used for the Brillouin zone integral. Using the second-order variational
procedure, we include the SOC \cite{socref}, which has been found to play an
important role in the $5d$ system. The self-consistent calculations are
considered to be converged when the difference in the total energy of the
crystal does not exceed 0.01 mRy. Despite the fact that the $5d$ orbitals
are spatially extended, recent theoretical and experimental work has given
evidence on the importance of Coulomb interactions in $5d$ compounds \cite%
{Rev-5d-1,Rev-5d-2,Rev-5d-3}. We utilize the LSDA + $U$ scheme \cite{LDA+U}
to take into account the effect of Coulomb repulsion in $5d$ orbital. We
vary the parameter $U$ between 2.0 and 3.0 eV and find that the essential
properties are independent on the value of $U$.

\begin{figure}[tbp]
\centering\includegraphics[width=0.45\textwidth]{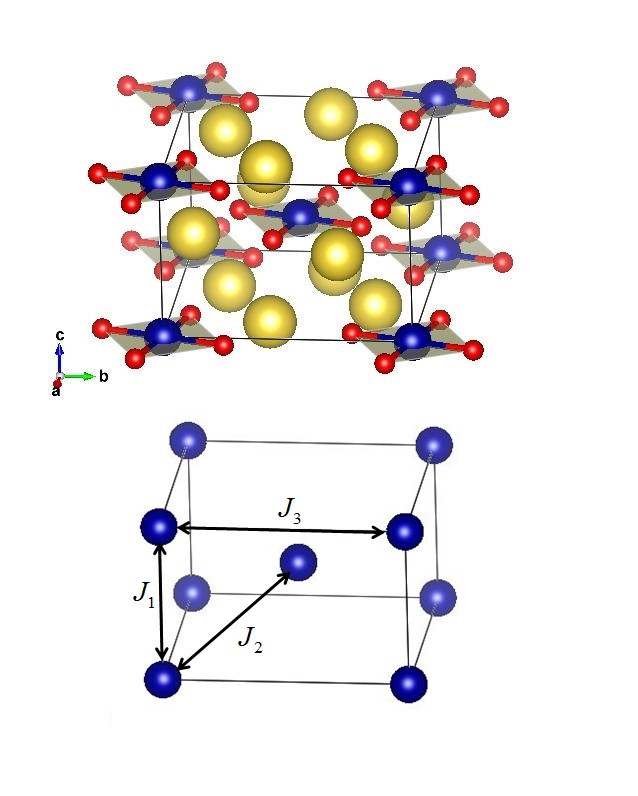}
\caption{Crystal structure of Na$_{4}$IrO$_{4}$. The yellow, blue, and red
balls represent the Na, Ir, and O ions, respectively. The nearest-neighbor,
next-nearest-neighbor and third-nearest-neighbor exchange interactions for
Ir magnetic moments are shown by $J_{1}$, $J_{2}$, $J_{3}$ respectively,
which are the parameters in the Heisenberg model $H=\sum_{i<j}J_{ij}S_{i}%
\cdot S_{j}.$}
\label{structure}
\end{figure}

As shown in Fig. \ref{structure}, Na$_{4}$IrO$_{4}$ crystallizes in the
tetragonal structure (space group I4/m) \cite{str}. The lattice constants of
Na$_{4}$IrO$_{4}$ are $a$ = 7.17 \AA\ and $c$ = 4.71 \AA \cite{str}. There
is only one formula unit in the primitive unit cell, and the nine atoms in
the unit cell are located at three nonequivalent crystallographic sites: Ir
atoms occupy the $2a$ position: (0,0,0), while both Na and O reside at the $%
8h$ sites: $(x,y,z)$\cite{str}. The square-planar IrO$_{4}$ oxoanion occurs
in the $ab$-plane, and is slightly rotated about the $c$ axis \cite{str}.
The Ir ions occupy the center of the square-planar coordination environment.
The average distance of four Ir-O bonds in the square-planar IrO$_{4}$ is
1.91 \AA , which is similar to the Ir-O bond length in IrO$_{6}$ octahedron.
Instead of the face (or edge, corner)-sharing structure of octahedrons, the
IrO$_{4}$ moiety is quite isolated as shown in Fig. \ref{structure}, thus
the Ir-Ir distance is quite large. These remarkable structural features
significantly affect the electronic structure and magnetic properties of Na$%
_{4}$IrO$_{4}$\ as shown in the following sections.

\section{Band structures}

\begin{figure*}[tbp]
\centering\includegraphics[width=18cm]{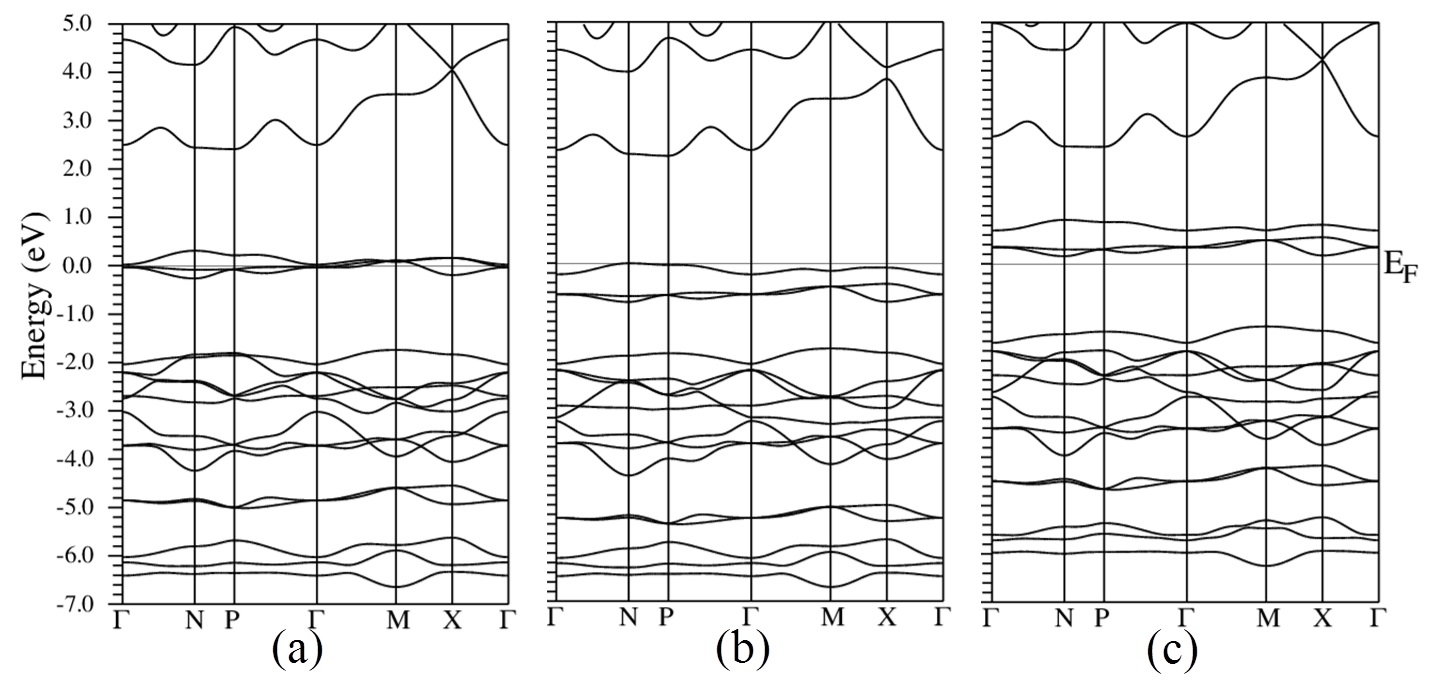}
\caption{Band structures of Na$_{4}$IrO$_{4}$. (a) represents the LDA
calculation while (b) and (c) represent spin-up and spin-down channel from
LSDA calculation with FM configuration. The Fermi energy is set to zero. }
\label{bandstr}
\end{figure*}

\begin{figure}[tbp]
\centering\includegraphics[width=0.45\textwidth]{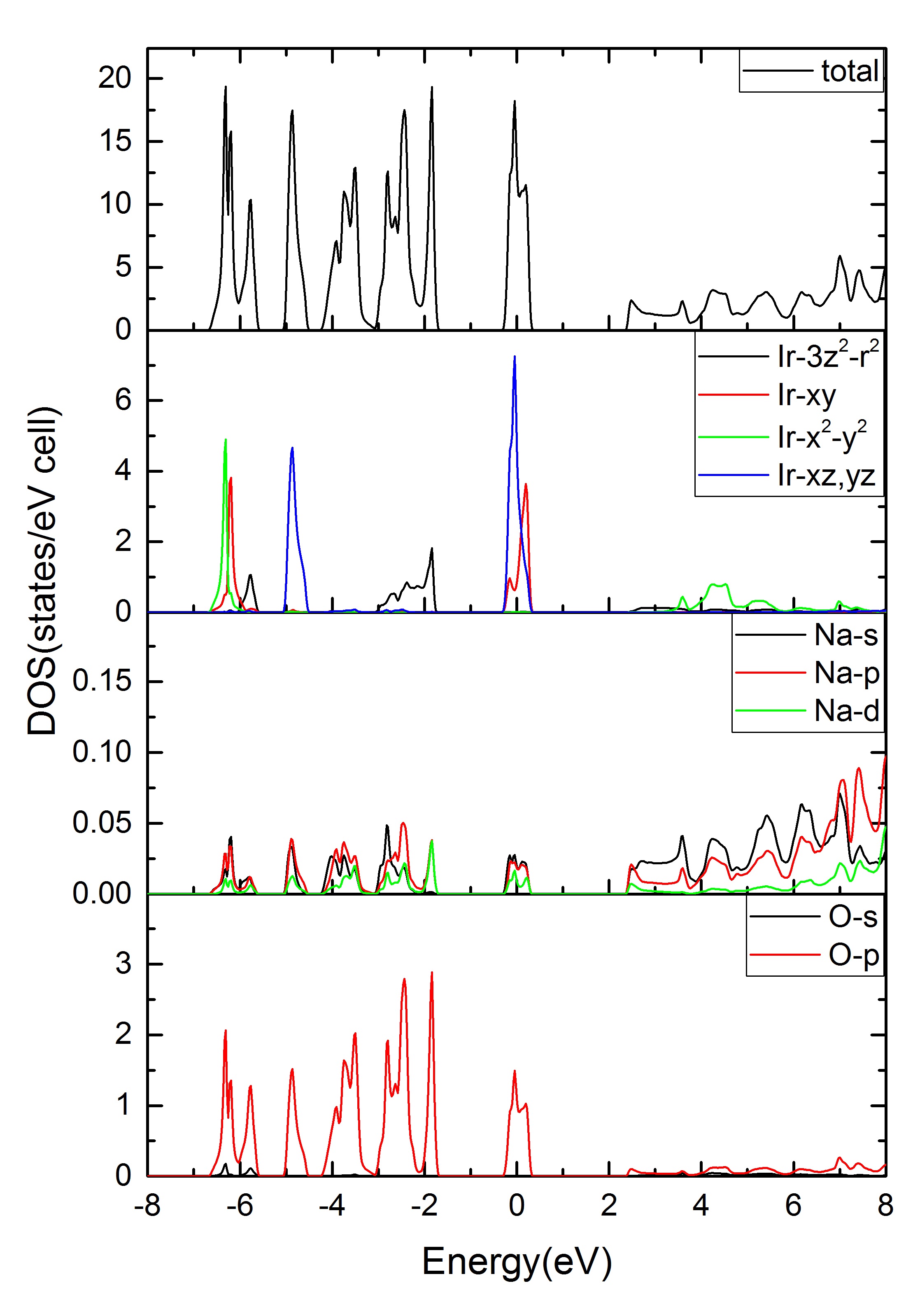}
\caption{Partial density of states (PDOS) of Na$_{4}$IrO$_{4}$ calculated by
the method of LDA calculation. The Fermi energy is set to zero. }
\label{dos}
\end{figure}

To clarify the basic electronic features, we perform nonmagnetic LDA
calculation, and show the band structures and the density of states (DOS) in
Fig. \ref{bandstr}(a) and Fig. \ref{dos}, respectively. As shown in Fig. \ref%
{dos}, that O-$2p$ states are mainly located between -7.0 and -1.0 eV while
the Na $3s$ and $3p$ bands appear mainly above 3.0 eV which is much higher
than the Fermi level and also appear between -7.0 and -1.0 eV contributed
mainly by the O-$2p$ states, indicating the non-negligible hybridization
between Na and O states despite that Na is highly ionic. Hence the chemical
valence for Na is $+1$\ while that for O is $-2$. As a result, the nominal
valence of Ir in Na$_{4}$IrO$_{4}$\ is +4, and the electronic configuration
of Ir ion is $5d^{5}$. It is well known that in the octahedral environment\
the $5d$ orbitals will split into the $t_{2g}$ and $e_{g}$ states, and the
strong SOC in $5d$ electrons splits the $t_{2g}$ states into $J_{eff}$ = 1/2
and $J_{eff}$ = 3/2 bands \cite{J-1/2 1,J-1/2 2}. Compared with the IrO$_{6}$
octahedra, the upper and lower O$^{2-}$ ions are absent in the square-planar
IrO$_{4}$ oxoanion. Consequently the Ir-5$d$ orbitals split into three
non-degenerate orbitals: $d_{3z^{2}-r^{2}},\ d_{xy},\ d_{x^{2}-y^{2}}$, and
doubly degenerate $d_{xz}/d_{yz}$ ones. There are in total 13 bands in the
energy range from -7.0 eV to -1.0 eV, as shown in Fig. \ref{bandstr}(a). The 
$d_{3z^{2}-r^{2}}$\ states appear mainly between -3.0 eV to -2.0 eV, while
the remaining 12 bands are contributed by O-$2p$ states. Mainly located
above 4 eV, $d_{x^{2}-y^{2}}$ states have also large distribution around
-6.5 eV due to the strong hybridization with O-$2p$ bands. The $%
d_{xz}/d_{yz} $ and $d_{xy}$\ orbitals are mainly located from -1.0 to 1.0
eV, while the $d_{xy}$ state is slightly higher in energy. As shown in Fig. %
\ref{bandstr}(a) and Fig. \ref{dos},\ these bands are separated from other
bands, and around the Fermi level the orbital splitting can be displayed by
the left panel of Fig. \ref{cfs}. The dispersion of the $5d$\ bands around
Fermi level is very narrow, due to that the IrO$_{4}$\ moiety is quite
isolated in the crystal structure. As shown in Fig. \ref{dos}, the DOS at
Fermi level is rather high, which indicates the magnetic instability.

\begin{figure}[tbp]
\centering\includegraphics[width=8cm]{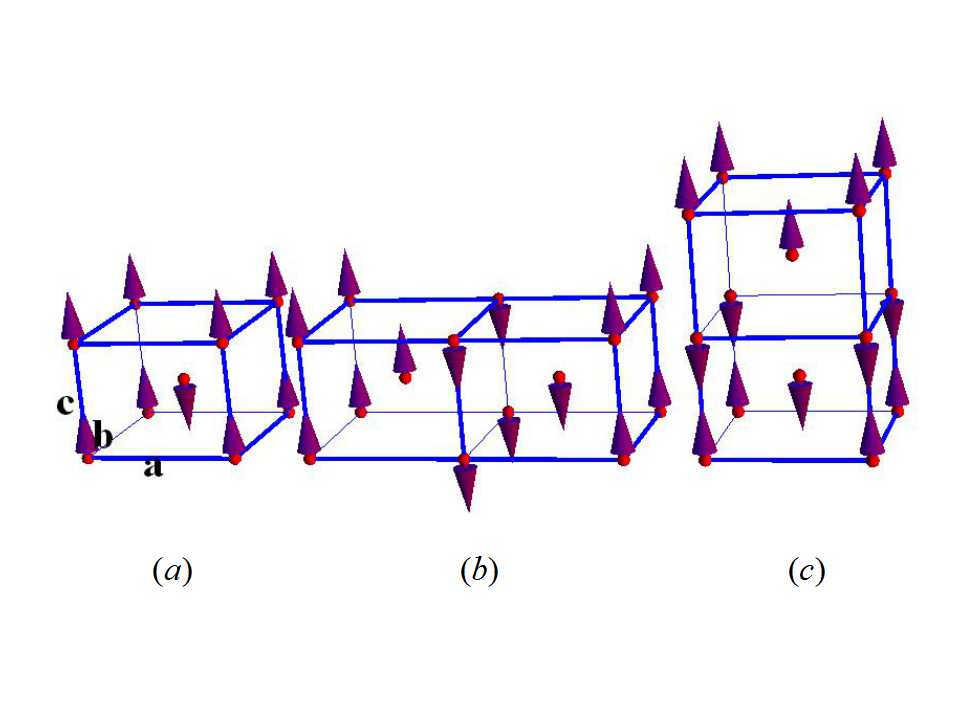}
\caption{The AFM magnetic configurations of Na$_{4}$IrO$_{4}$ which we
considered in DFT calculations. For clarity only Ir atoms are shown. (a),
(b) and (c) represent AFM-1, AFM-2 and AFM-3 configurations, respecitvely.}
\label{magneticstructure}
\end{figure}

To understand the magnetic properties, we also perform a spin polarized
calculation and show the band structures of ferromagnetic (FM) configuration
in Fig. \ref{bandstr}(b) and \ref{bandstr}(c). Basically, the $%
d_{3z^{2}-r^{2}}$ states are fully occupied while the $d_{x^{2}-y^{2}}$ ones
are empty, and the spin polarization has a relatively small effect on these
bands. On the other hand, the partially occupied $d_{xz}/d_{yz}$ and $d_{xy}$
states are significantly affected, and these bands have about 1 eV exchange
splitting, as shown in Fig. \ref{bandstr}(b) and \ref{bandstr}(c). LSDA
calculation for FM configuration gives a insulating solution with a band gap
of 0.16 eV. Experiment reveals that Na$_{4}$IrO$_{4}$ has a long-range
antiferromagnetic (AFM) order at low temperature \cite{414}. Thus we also
explore the magnetic configuration. In addition to the FM configuration, we
also consider three AFM states: AFM-1 where Ir atoms at the body center and
corners have opposite spin orientations, AFM-2 where Ir atoms couple
anti-ferromagnetically along a-axis, AFM-3 where Ir atoms couple
anti-ferromagnetically along c-axis (See Fig. \ref{magneticstructure} for
the magnetic structures of different AFM\ configurations). The relative
total energies and magnetic moments for the four magnetic configurations are
summarized in Table. \ref{afm123}. Different magnetic configurations have
similar calculated magnetic moments. This indicates that the magnetism in Na$%
_{4}$IrO$_{4}$ is quite localized. The distance between IrO$_{4}$ oxoanion
is quite large as shown in Fig. \ref{structure}, thus the effective hopping
between Ir ions in Na$_{4}$IrO$_{4}$ is very weak. As a result, the
magnetism in Na$_{4}$IrO$_{4}$\ is very localized and different magnetic
configurations have only small effects on the band structures. Regardless of
the magnetic configuration, our numerical results show that the $5d$\
electronic configuration can always be described by\textbf{\ }$%
d_{3z^{2}-r^{2},\uparrow }^{1}d_{3z^{2}-r^{2},\downarrow }^{1}d_{xz,\uparrow
}^{1}d_{yz,\uparrow }^{1}d_{xy,\uparrow }^{1}$. While for most of $5d$
transition metal oxides, the magnetization is quite itinerant and the
magnetic configuration strongly affect the band structure \cite{Rev-5d-3}.
The calculated magnetic moment at Ir site is around 1.35 $\mu _{B}$,
considerably small than that of $S$ = $3/2$ configuration. Due to the strong
hybridization between Ir-$5d$ and O-$2p$ states, there is also considerable
magnetic moment located at O site. As shown in Table \ref{afm123}, the AFM-1
configuration has the lowest total energy. Although we only consider four
magnetic configurations, we believe that the AFM-1 is indeed the magnetic
ground state configuration as discussed in the following sections.

\begin{table}[tbp]
\caption{The calculated total energy (in meV) per unit cell and magnetic
moments (in $\protect\mu $B) for the four magnetic configurations from LSDA
and LSDA + $U$ ($U$ = 2 eV) calculations. The total energy of AFM1 state is
set to zero.}
\label{afm123}%
\begin{tabular}{lllll|llll}
\hline\hline
& \multicolumn{4}{c|}{LSDA} & \multicolumn{4}{|c}{LSDA+$U$} \\ 
& FM & AFM1 & AFM2 & AFM3 & FM & AFM1 & AFM2 & AFM3 \\ \hline
E$_{total}$ & 44.5 & 0 & 19.7 & 17.9 & 22.8 & 0 & 10.8 & 8.4 \\ 
m$_{Ir}$ & 1.40 & 1.32 & 1.34 & 1.35 & 1.48 & 1.45 & 1.46 & 1.46 \\ 
m$_{O}$ & 0.27 & 0.25 & 0.26 & 0.26 & 0.26 & 0.25 & 0.26 & 0.26 \\ 
\hline\hline
\end{tabular}%
\end{table}

\begin{figure}[tbp]
\centering\includegraphics[width=0.45\textwidth]{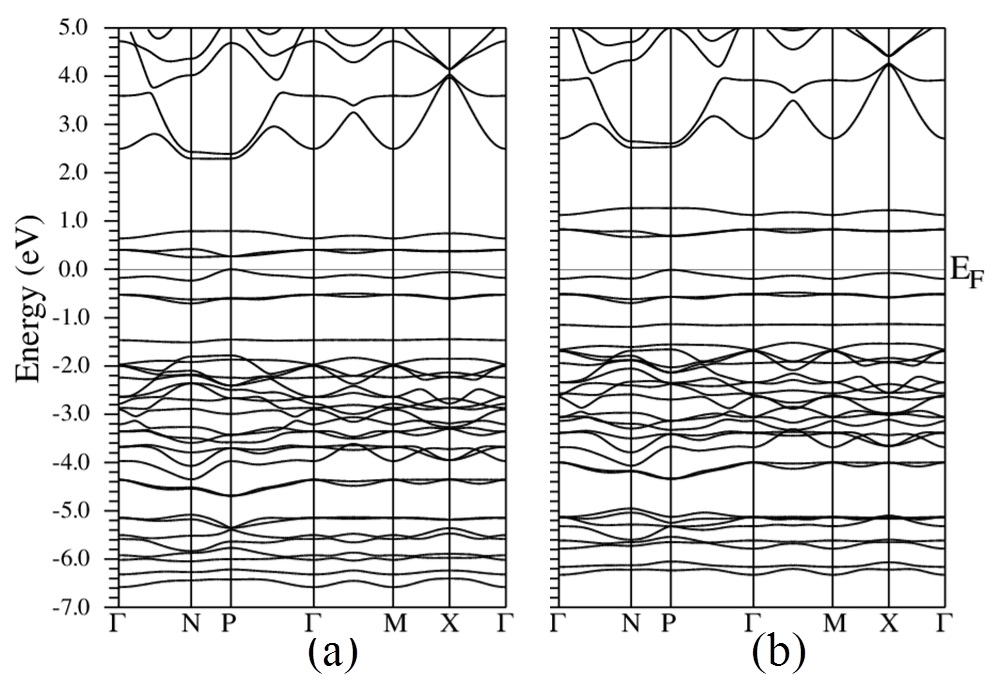}
\caption{Band structure of AFM-1 configuration calculated by (a) LSDA (b)
LSDA\ +\ $U$ (= 2 eV) method.}
\label{afm1-band}
\end{figure}

\begin{figure}[tbp]
\centering\includegraphics[width=0.45\textwidth]{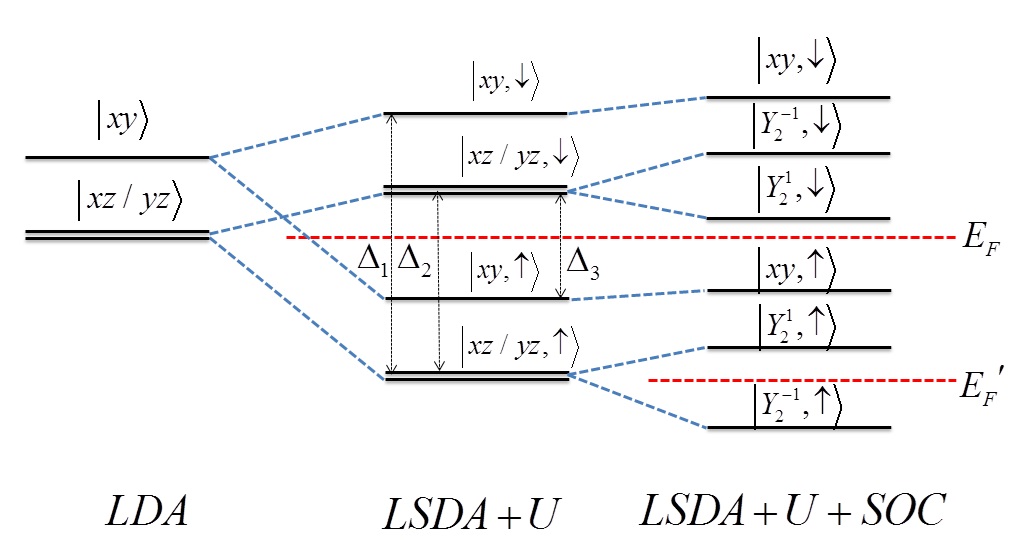}
\caption{Schematic picture of orbital occupation around Fermi level. From
LDA (left panel), LSDA+$U$ (middle panel) and LSDA+$U$+SOC (right panel)
calculations. LSDA calculation has the same pattern with LSDA+$U$
calculations and is not presented here. $E_{F}$ represent the Fermi-level of
Na$_{4}$IrO$_{4}$. As mentioned in the maintext, substituting Ir ions by Re
ions, the Fermi-level shifts to $E_{F}^{\prime }$, which significantly
enhances the MAE.}
\label{cfs}
\end{figure}

As the importance of electronic correlation for $5d$ orbitals has been
recently emphasized \cite{Rev-5d-1,Rev-5d-2,Rev-5d-3}, we utilize the LSDA + 
$U$ scheme, which is adequate for the magnetically ordered insulating ground
states, to consider the electronic correlation in Ir-$5d$ states. The
estimates for the values of $U$ have been recently obtained between 1.4 and
2.4 eV in layered Sr$_{\mathbf{2}}$IrO$_{\mathbf{4}}$\ and Ba$_{2}$IrO$_{4}$ 
\cite{5dUvalue}. The Ir ion in the IrO$_{4}$\ moiety has only four nearest
neighbors. Moreover IrO$_{4}$\ moieties are highly-isolated, thus we
generally expect that the value of $U$ in Na$_{4}$IrO$_{4}$\ is larger than
that in other $5d$ transition metal oxides. We have varied the value of $U$\
from 2.0 to 3.0 eV, the electronic structure and magnetic properties depend
moderately on $U$\ and the numerical calculations show that the essential
properties and our conclusions do not depend on the value of $U$. Thus we
only show the results with $U$ = 2 eV at follows. Similarly we consider the
four magnetic configurations and the results of relative total energies and
magnetic moments are also summarized in Table. \ref{afm123} while AFM-1
state still has the lowest energy. Including $U$ will enhance the exchange
splitting in $5d$ bands, and slightly enlarge the calculated magnetic
moments as shown in Table \ref{afm123}. The band structures of Na$_{4}$IrO$%
_{4}$ with AFM-1 order from the LSDA + $U$ calculation is presented in Fig. %
\ref{afm1-band}. The result from the LSDA calculation within the AFM-1
configuration is also shown for comparison. Compared with the LSDA
calculation, the bands within LSDA + $U$ scheme are narrower but the order
of crystal field splitting pattern and electronic occupation does not
change. The LSDA + $U$ calculation predicts a little bigger magnetic moment
for Ir ions (1.45 $\mu _{B}$) and a larger gap of about 0.57 eV, as shown in
Table. \ref{afm123} and Fig. \ref{afm1-band}. We also show the $d$ orbital
splitting under the crystal field of square plane in the middle panel of
Fig. \ref{cfs}, and the electronic occupation pattern is decided by the
competition between the crystal field splitting and Hund's rule. It is worth
mentioning that the results such as the crystal splitting pattern, are not
dependent on the value of $U$.

\begin{figure}[tbp]
\centering\includegraphics[width=0.45\textwidth]{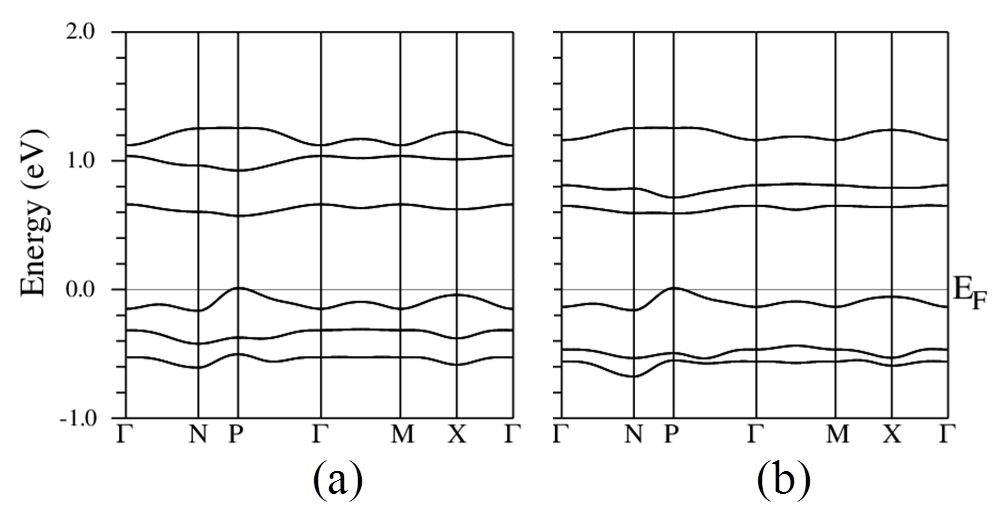}
\caption{Band structure of Na$_{4}$IrO$_{4}$ calculated by the method of
LSDA + $U$ + SOC ($U$ = 2 eV) calculation. (a) and (b) represent spin
orientations along (001) direction and (100) direction, respectively.}
\label{socband}
\end{figure}

The strong SOC in $5d$ atoms usually significantly affects the band
dispersions, thus we also perform the LSDA + $U$ + SOC calculations. Since
the IrO$_{4}$ moiety is in the $ab$-plane, we perform the LSDA + $U$ + SOC
calculations with spin orientations perpendicular to $ab$-plane and lying in 
$ab$-plane, i.e. the spin orientations are along (001), (010) and (100)
directions. Our calculations show that the (001) is the easy axis and (100)
is the hard axis. We list the calculated magnetic moments and total energies
in Table \ref{afm123-soc} and find that AFM-1 state is still the ground
state configuration. Unlike LSDA + $U$ calculations, the degeneracy of $%
d_{xz}$ and $d_{yz}$ is removed by SOC, as shown in Fig. \ref{socband} and
the the right panel of Fig. \ref{cfs}. The most remarkable feature is the
huge MAE. The MAE of Na$_{4}$IrO$_{4}$ is around 12 meV per Ir atom with the
highly preferential easy axis being out of ab-plane. It is easy to see from
Table. \ref{afm123-soc} that for all of the four magnetic configurations,
(001) direction is the easy axis and the MAE have the similar values from
11.6 to 12.6 meV per Ir atom.

Large magnetic anisotropy energy (MAE) is desirable for magnetic devices.
Recently, there has been considerable research interest in studying
materials with a large MAE. Most of them are two dimensional materials or
adatoms on surfaces. For example, Co atoms deposited on a Pt (111) surface 
\cite{pt}, Fe or Mn atoms absorbed on the CuN surface \cite{cun}, and Co or
Fe atoms on Pd or Rh (111) surface \cite{pdrh}. In addition, Rau et al. \cite%
{mgo} found a giant MAE for the Co atoms absorbed on top of the O sites of
MgO (001) surface. Generally, the bulk materials exhibit relatively small
MAE of a few $\mu eV$ \cite{uev1,uev2} while anisotropy energies are larger
by about three orders of magnitude for multilayers and surface systems \cite%
{uev2}. MAE originates from the interaction of the atom's orbital magnetic
moment and spin angular moment, thus an important factor of MAE is the
strength of SOC, which increases from $3d$ to $5d$ metals. Another important
factor is the special coordination environment, since a ligand field often
quenches the orbital moment. Since the IrO$_{4}$\ in Na$_{4}$IrO$_{4}$\
shows an isolated planar structure, large MAE is expected.

In order to confirm the giant value of MAE, we also calculate the variation
of the total energy by changing the magnetization direction with the\textbf{%
\ }force theorem. In this case, there is no need to converge a complete
self-consistency cycle. The evaluated MAE using force theorem gives the
similar values. We try to understand the magnetic properties in following
sections.

\begin{table}[tbp]
\caption{The calculated total energy (in meV) per unit cell and magnetic
moments (in $\protect\mu $B) for the four magnetic configurations from LSDA
+ SOC + $U$ ($U$ = 2 eV) calculations with (001) and (100) spin
orientations. The total energy of AFM-1 state with (001) magnetization
direction is set to zero. MAE (in meV) per Ir atom for four magnetic
configurations are also summarized in the table.}
\label{afm123-soc}%
\begin{tabular}{lll|ll|ll|ll}
\hline\hline
& \multicolumn{2}{c|}{FM} & \multicolumn{2}{|c|}{AFM-1} & 
\multicolumn{2}{|c|}{AFM-2} & \multicolumn{2}{|c}{AFM-3} \\ 
& (001) & (100) & (001) & (100) & (001) & (100) & (001) & (100) \\ \hline
E$_{total}$ & 22.3 & 34.4 & 0 & 11.6 & 11.1 & 22.8 & 8.9 & 21.5 \\ 
m$_{Ir}$(spin) & 1.37 & 1.38 & 1.34 & 1.35 & 1.36 & 1.36 & 1.36 & 1.36 \\ 
m$_{Ir}$(orbital) & 0.11 & 0.10 & 0.10 & 0.10 & 0.10 & 0.10 & 0.10 & 0.10 \\ 
m$_{O}$ & 0.25 & 0.24 & 0.24 & 0.23 & 0.24 & 0.24 & 0.24 & 0.24 \\ \hline
MAE & \multicolumn{2}{c|}{12.1} & \multicolumn{2}{|c|}{11.6} & 
\multicolumn{2}{|c|}{11.7} & \multicolumn{2}{|c}{12.6} \\ \hline\hline
\end{tabular}%
\end{table}

\section{Spin Model}

As shown in Table \ref{afm123}, the calculated magnetic moments for the
different magnetic configurations are similar, thus the total energy
differences between the different magnetic configurations are mainly
contributed by the inter-atomic exchange interaction where SOC is not
considered. This allows us to estimate the exchange couplings by the
energy-mapping analysis (see Appendix). As shown in Fig. \ref{structure}, we
consider three spin exchange paths. $J_{1}$ is the nearest-neighbor Ir-Ir
exchange coupling along c-axis, $J_{2}$ is the next-nearest-neighbor one
along diagonal line while $J_{3}$ is the 3rd-nearest-neighbor one along $a/b$%
-axis. The distance in $J_{3}$ (7.1 $\mathring{A}$) is much longer than $%
J_{1}$ (4.7 $\mathring{A}$) and $J_{2}$ (5.6 $\mathring{A}$). Compared with
LSDA + $U$, LSDA overestimates the hopping, consequently gives larger
exchange parameters as shown in Table \ref{Jvalue}. $J_{1}$, $J_{2}$ and $%
J_{3}$ are all AFM. Although $J_{1}$ has the nearest-neighbor exchange path, 
$d_{3z^{2}-r^{2}}$ orbital is fully occupied in both up and down spin
channel as mentioned in the previous section and the hoppings for the other $%
d$ orbitals are relatively small. Therefore, it is easy to understand that
the value of $J_{1}$ is less than $J_{2}$. Thus $J_{2}$ dominates over the
others in strength, while $J_{3}$ is nearly negligible due to the much long
distance, as shown in Table \ref{Jvalue}. Although the spin exchange
couplings $J_{1}$-$J_{3}$ decrease with increasing $U$ values, $J_{2}$ is
always dominated while $J_{3}$ is nearly negligible.\ The exchange
interaction in magnetic insulators is predominantly caused by the so-called
superexchange --- which is due to the overlap of the localized orbitals of
the magnetic electrons with those of intermediate ligands. The Ir-Ir
distance in Na$_{\mathbf{4}}$IrO$_{\mathbf{4}}$ is very large, thus the
value of exchange interaction $J$\ with longer distance should be very
smaller and have no influence on the magnetic ground state. Thus we believe
that the strongest $J_{2}$\ makes AFM-1 as the ground state, in agreement
with the total energy calculations.

Based on the $J_{1}$-$J_{3}$ parameters from LSDA + $U$, we calculate the
Curie-Weiss temperature $\theta $ and N\'{e}el temperature $T_{N}$ using the
mean-field approximation theory\cite{theta}. $\theta $ is estimated to -105
K while $T_{N}$ is about 56 K. The values of -105 K and 56 K are both
somewhat larger but qualitatively consistent with the experimental ones of
-78 K and 25 K, respectively. Since the mean-field approximation theory
often overestimate the Curie-Weiss and N\'{e}el temperatures, our mapping $%
J_{1}$-$J_{3}$ parameters are thought to be in reasonable agreement with
experimental results.

\section{Magnetic anisotropy energy}

In order to understand the origin of the giant MAE, we start from a
generalized symmetry allowed spin model of Na$_{4}$IrO$_{4}$ (See Appendix):

\begin{equation}
H_{S}=-K\sum_{i}{S_{i}^{z}}^{2}+\frac{1}{2}\sum_{<i,j>,\alpha \beta
}J_{ij}^{\alpha \beta }S_{i}^{\alpha }S_{j}^{\beta },  \label{J}
\end{equation}%
where the first term represents the single ion anisotropy Hamiltonian, the
second one is the inter-atomic exchange Hamiltonian, $i,j$ label the Ir ions
and $\alpha ,\beta $ take $x,y,z$. Due to the inversion symmetry, $%
J_{ij}^{\alpha \beta }=J_{ij}^{\beta \alpha }$, which means there is no
Dzyaloshinskii-Moriya interaction \cite{D,M}. We only consider the exchange
neighbors $<ij>^{\prime }s$ to the 3rd nearest-neighbor, which are denoted
by $J_{1},J_{2},J_{3}$ in order as shown in Fig. \ref{structure}. For $J_{1}$%
, due to the $C_{4}$ rotation symmetry, $J_{1}^{\alpha \beta }=\delta
_{\alpha \beta }J_{1}^{\alpha \alpha }$ and $J_{1}^{xx}=J_{1}^{yy}$. While
for $J_{2}$\ and $J_{3}$, the non-diagonal terms (i.e. $J^{xy},J^{xz}$\ and $%
J^{yz}$) is symmetry-allowed, however these terms are proportional to the
product of $\lambda ^{2}$\ and isotropic exchange \cite{nondiagonal}, and
should be very small, thus we ignore them hereafter.

Using the similar energy-mapping method (See Appendix), we estimate the
parameters in Eq. (\ref{J}) and show the results in Table. \ref{Jvalue}. It
is clear that the anisotropy of spin exchange is small, especially for the
dominating spin exchange, where $J_{2}$ shows a small difference between $%
J_{2}^{xx}$, $J_{2}^{yy}$\ and $J_{2}^{zz}$. The different spin
configurations have almost the same value of MAE, and the anisotropy of spin
exchange coupling parameters is little, indicating that MAE is dominated by
the single-ion anisotropy.

\begin{table}[tbp]
\caption{Isotropic spin exchange parameters (in meV) and anisotropic spin
exchange parameters evaluated by energy-mapping analysis from LSDA, LSDA + $%
U $, LSDA + SOC + $U$ ($U$ = 2 eV) calculations, respectively.}
\label{Jvalue}%
\begin{tabular}{ll|l|lll}
\hline\hline
& LSDA & LSDA+$U$ & \multicolumn{3}{|l}{LSDA+$U$+SOC} \\ 
&  &  & $J_{i}^{xx}$ & $J_{i}^{yy}$ & $J_{i}^{zz}$ \\ \hline
$J_{1}$/meV & \multicolumn{1}{c|}{0.97} & \multicolumn{1}{|c|}{0.66} & 0.32
& 0.32 & 0.51 \\ 
$J_{2}$/meV & \multicolumn{1}{c|}{2.47} & \multicolumn{1}{|c|}{1.27} & 1.21
& 1.35 & 1.24 \\ 
$J_{3}$/meV & \multicolumn{1}{c|}{0.56} & \multicolumn{1}{|c|}{0.14} & 0.02
& 0.02 & 0.01 \\ 
$K$/meV & \multicolumn{1}{c|}{-} & \multicolumn{1}{|c|}{-} & 
\multicolumn{3}{|c}{5.4} \\ \hline\hline
\end{tabular}%
\end{table}

To understand the origin of single-ion anisotropy, we consider the crystal
field splitting, electronic occupation shown in Fig. \ref{cfs}, and the SOC
Hamiltonian $\lambda L\cdot S$ where $\lambda $ is the SOC constant. With
the spin direction described by the two angles $(\theta ,\varphi )$, where $%
\theta $ and $\varphi $ are the azimuthal and polar angles of the spin
orientation with respect to the local coordinate environment, the $\lambda
L\cdot S$ term can be written as \cite{soc}

\begin{align}
H_{so}& =\lambda \hat{S}_{z^{\prime }}(\hat{L}_{z}\cos \theta +\frac{1}{2}%
\hat{L}_{+}e^{-i\varphi }\sin \theta +\frac{1}{2}\hat{L}_{-}e^{i\varphi
}\sin \theta )  \notag \\
& \quad +\frac{\lambda }{2}\hat{S}_{+^{\prime }}(-\hat{L}_{z}\sin \theta -%
\hat{L}_{+}e^{-i\varphi }\sin ^{2}\frac{\theta }{2}+\hat{L}_{-}e^{i\varphi
}\cos ^{2}\frac{\theta }{2})  \notag \\
& \quad +\frac{\lambda }{2}\hat{S}_{-^{\prime }}(-\hat{L}_{z}\sin \theta +%
\hat{L}_{+}e^{-i\varphi }\cos ^{2}\frac{\theta }{2}+\hat{L}_{-}e^{i\varphi
}\sin ^{2}\frac{\theta }{2})
\end{align}

Using perturbation theory by treating the SOC Hamiltonian as the
perturbation combined with the $d$ orbital occupation pattern, we can get
the associated energy lowering:

\begin{eqnarray}
\Delta E^{(1)} &=&\sum_{i}\left\langle i\left\vert H_{so}\right\vert
i\right\rangle  \notag \\
\Delta E^{(2)} &=&-\sum\limits_{i,j}\frac{\left\vert \left\langle
i\left\vert H_{so}\right\vert j\right\rangle \right\vert ^{2}}{\left\vert
e_{i}-e_{j}\right\vert }
\end{eqnarray}

where $i$ represents an occupied $d$-level state with energy $e_{i}$ while $%
j $ represents an unoccupied $d$-level state with energy $e_{j}$, and the
third and higher order perturbations are not given here. In Na$_{4}$IrO$_{4}$%
, where the SOC has not been considered, $d_{xz,\uparrow }$ and $%
d_{yz,\uparrow }$ are doubly-degenerate. We can see that the degeneracy of $%
d_{xz,\uparrow }$ and $d_{yz,\uparrow }$ is lifted by the SOC and they split
to $\left\vert Y_{2}^{1},\uparrow \right\rangle $ and $\left\vert
Y_{2}^{-1},\uparrow \right\rangle $. The splitting is $\pm \frac{\lambda }{2}%
\left\vert \cos \theta \right\vert $\ according to the first order
perturbation, thus the splitting for the spin polarization of (001)
direction is larger than that for (100) direction, as shown in Fig. \ref%
{socband}. However, the $d_{xy,\uparrow }$, $d_{xz,\uparrow }$\ and $%
d_{yz,\uparrow }$\ are fully occupied and the band gap is quite big with
respect to the SOC constant $\lambda $. Thus the first order perturbation is
negligible and has no contribution to the single-ion anisotropy.

Therefore, we consider the second order perturbation. Note that in the
common $5d$ transition metal oxides with face (or edge, corner)-sharing
structure of oxygen octahedrons, the widths of the $t_{2g}$-block bandwidths
are relatively large while the $\left\vert e_{i}-e_{j}\right\vert $ values
are relatively small, so the perturbation theory does not lead to an
accurate estimation of MAE. But in Na$_{4}$IrO$_{4}$, the widths of the
bands around Fermi-level are about 0.2 eV and the $\left\vert
e_{i}-e_{j}\right\vert $\ value is around $\sim $2 eV, thus one can get the
quantitative value of MAE more accurately by the second order perturbation

\begin{equation}  \label{soc}
E= -\lambda ^{2}\cos ^{2}\theta \lbrack \frac{1}{4\Delta _{1}}+\frac{1}{%
4\Delta _{3}}-\frac{1}{2\Delta _{2}}]
\end{equation}

Here $\Delta _{i}$ is the splitting of on-site $d$-orbital energy levels, as
shown in Fig. \ref{cfs}. The orbitals of $d_{x^{2}-y^{2}}$ and $d_{z^{2}}$\
are far away from the Fermi-level and can be ignored. From the LSDA + $U$ ($%
U $ = 2 eV) calculations, we estimate the orbital energy levels by the
weight-center positions of DOS and get the values of $\Delta _{i}$. The
calculated values of $\Delta _{1}$,$\ \Delta _{2}$ and $\Delta _{3}$ are
1.82, 1.35 and 0.89 eV, respectively. Using these value of $\Delta _{i}$ and 
$\lambda $ = 0.5 eV, we get the MAE as 12.0 meV, consistent with the value
directly from DFT theory. Thus the second order perturbation is dominant in
MAE and higher order perturbations are believed to be little. One reason of
the giant MAE is the SOC strength, which is very strong for $5d$ electrons.
It is about 3 times as high as $4d$ electrons and an order of magnitude
higher than $3d$ electrons. Besides the strong SOC strength, the $d$-level
splitting is also a significant cause of the giant MAE. The $d$-level
splitting condition comes from the special square planar local environment
which indicates a giant anisotropy between in-plane and out-of-plane. The
long distances of IrO$_{4}$\ moieties make the strong local magnetization.
These factors together make the giant value of MAE.

\begin{figure}[tbp]
\centering\includegraphics[width=0.45\textwidth]{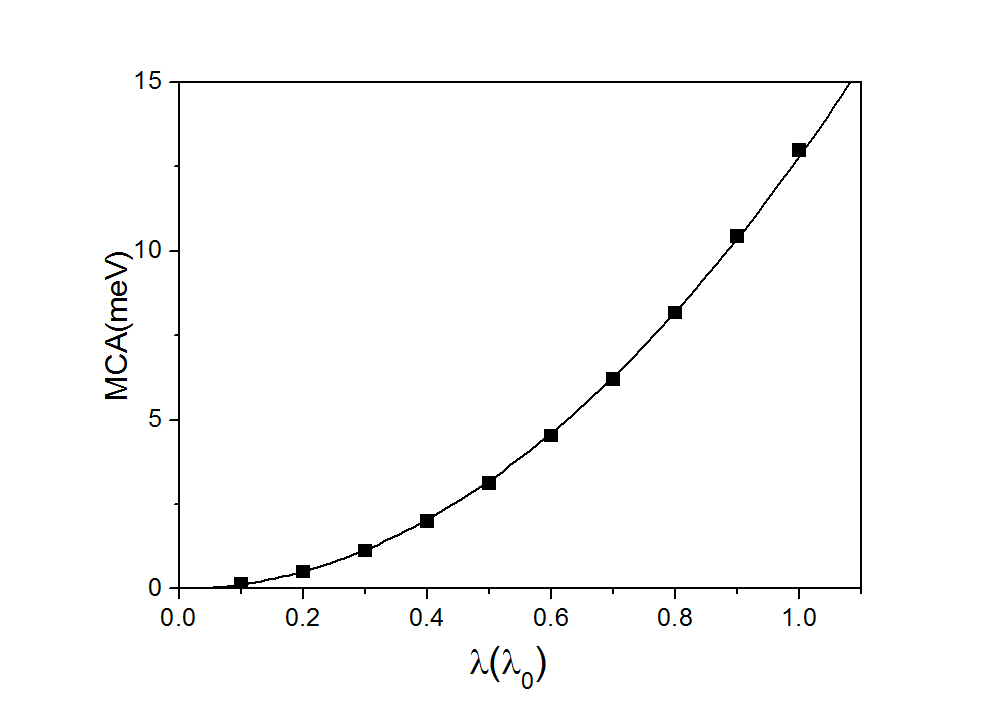}
\caption{SOC strength dependence of the magnetic anisotropy energy of Na$%
_{4} $IrO$_{4}$ calculated by the method of LSDA + SOC + $U$ ($U$ = 2 eV). }
\label{varysoc}
\end{figure}

We also calculate MAE with varying the SOC strength $\lambda $ within LSDA + 
$U$ + SOC scheme. As shown in Fig. \ref{varysoc}, it is obviously that the
MAE of Na$_{4}$IrO$_{4}$ is nearly proportional to the square of $\lambda $,
in accordance with Eqn. (\ref{soc}). 

\begin{figure}[tbp]
\centering\includegraphics[width=0.45\textwidth]{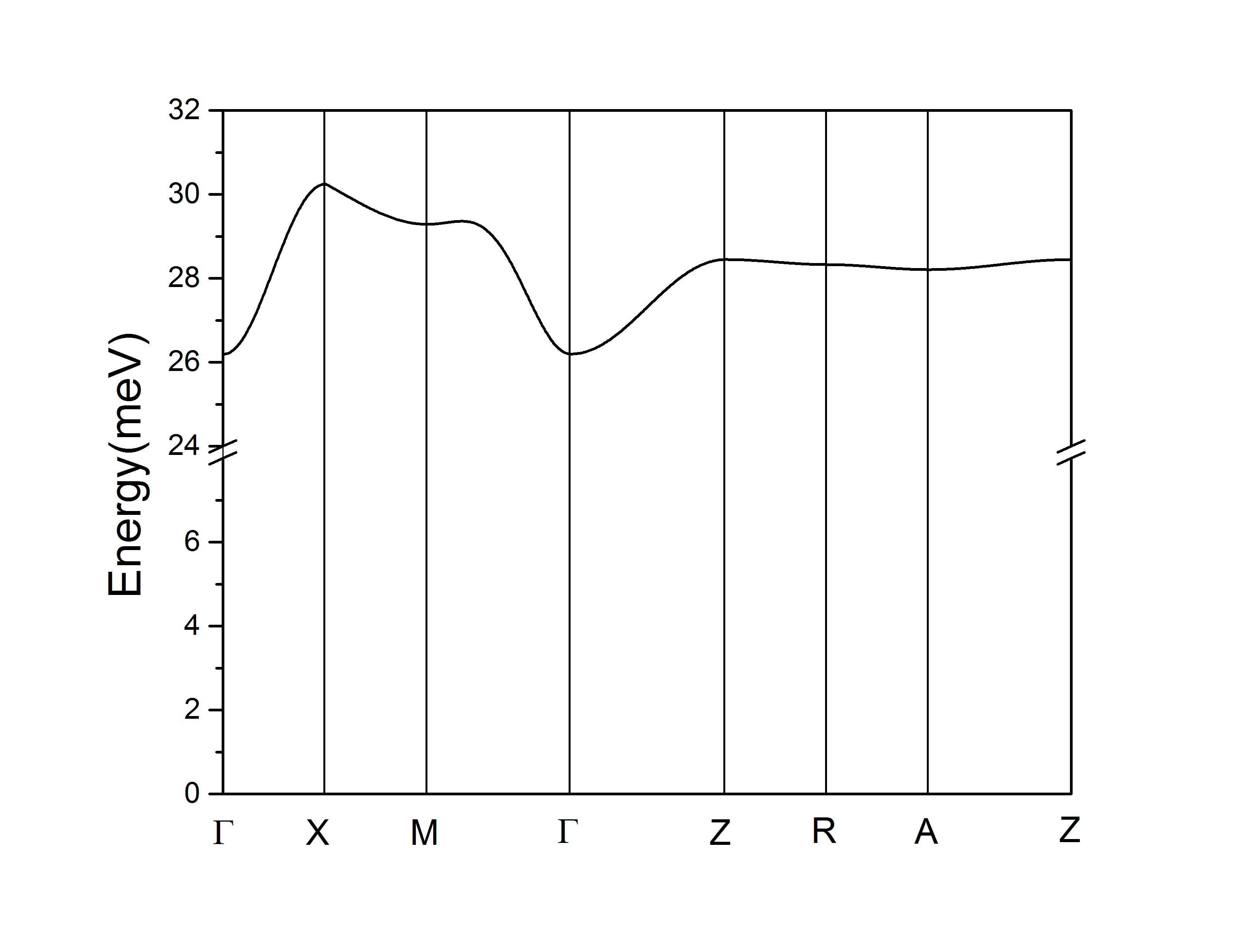}
\caption{Calculated spin-wave dispersion curves along high-symmetry axis for
the Na$_{4}$IrO$_{4}$.}
\label{spinwave}
\end{figure}

Using the calculated spin model parameters, one can obtain the magnon
spectrum on the basis of the Holstein-Prinmakoff transformation and the
Fourier transformation. We calculate the spin-wave dispersion along
high-symmetry axis and display the result in Fig. \ref{spinwave}. As shown
in Fig. \ref{spinwave}, there is a large spin gap of about 26 meV while the
width of spin-wave dispersion is only 4 meV. This is due to the large
single-ion anisotropy and relatively small exchange couplings.

\section{Material Design}

As shown in Fig. \ref{cfs}, for Na$_{4}$IrO$_{4}$, the exchange splitting is
large and there is a relatively big gap between occupied and unoccupied
states. Therefore the first order perturbation of SOC is very small. We
expect that if the Fermi-level shifts to the position of $E_{F}^{\prime }$\
as shown in Fig. \ref{cfs}, there is nonzero first order term and the MAE
will be enhanced significantly. We try to realize the Fermi-level shift
through substituting Ir ions in Na$_{4}$IrO$_{4}$\ by Re ions. The MAE of Na$%
_{4}$ReO$_{4}$ may be even larger and it may reach the limit of MAE even in
bulk materials, with the same size as Co or other atoms absorbed on top of
the O sites of MgO (001) surface \cite{3dmca,wuhua}.

\begin{figure}[tbp]
\centering\includegraphics[width=0.45\textwidth]{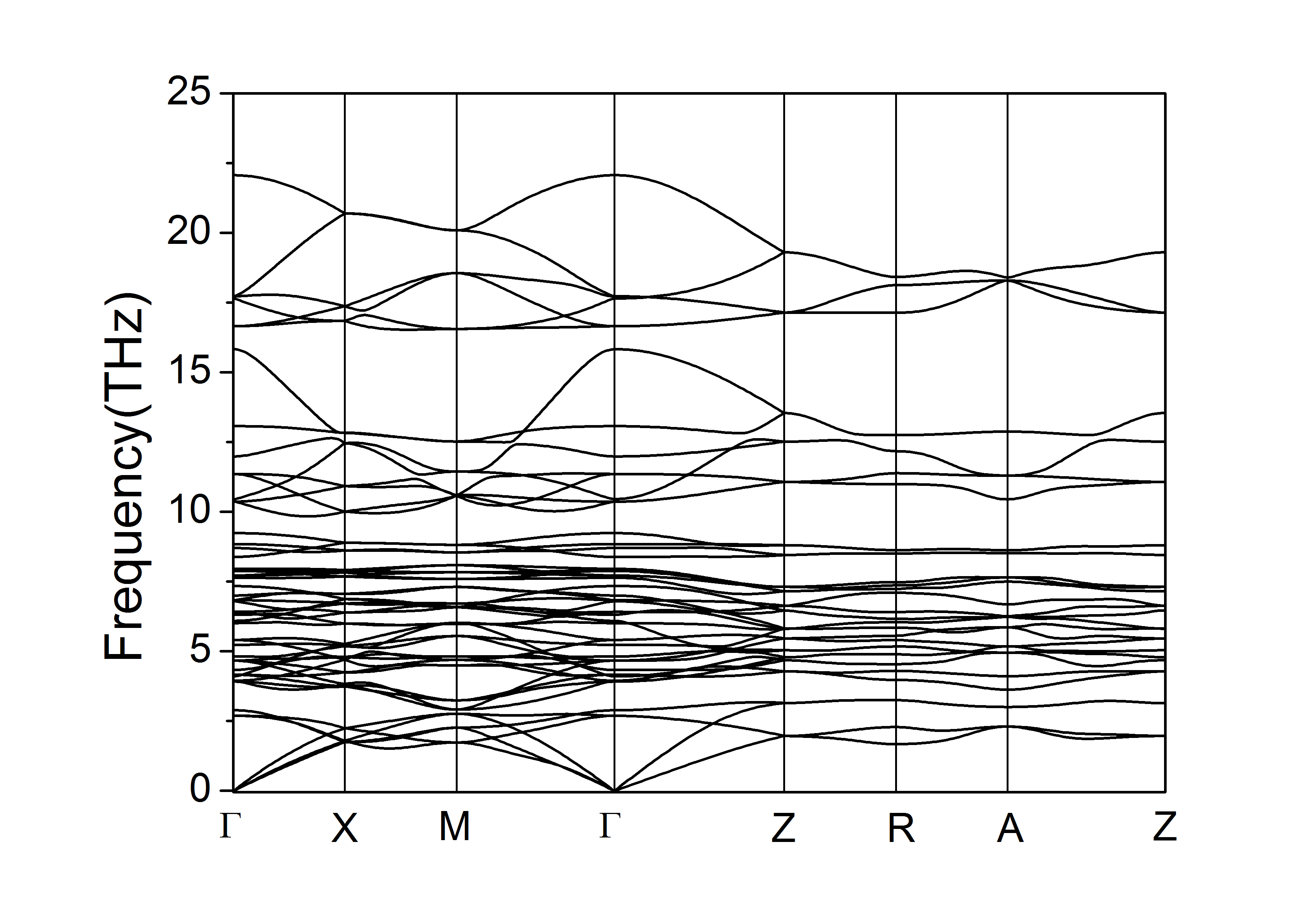}
\caption{Calculated phonon dispersion for Na$_{4}$ReO$_{4}$.}
\label{phonon}
\end{figure}

In order to examine the dynamic stability, we calculate phonon spectrum of Na%
$_{4}$ReO$_{4}$ (See Appendix), and show the calculated phonon spectrum
along high-symmetry lines in Fig. \ref{phonon}. All the phonon modes of Na$%
_{4}$ReO$_{4}$ are positive, indicating the structure is dynamically stable.

The calculated value of MAE is about 140 meV per Ir atom. It can be
explained by the same method using perturbation theory, where Na$_{4}$ReO$%
_{4}$ have two less occupied electrons. For Na$_{4}$ReO$_{4}$, with the
absence of SOC interaction, the orbitals of $d_{xz,\uparrow }$ and $%
d_{yz,\uparrow }$ are doubly-degenerate and half-occupied. With the presence
of SOC, the doubly-degenerate $d_{xz,\uparrow }$/$d_{yz,\uparrow }$ bands
split to $\left\vert Y_{2}^{1},\uparrow \right\rangle $ and $\left\vert
Y_{2}^{-1},\uparrow \right\rangle $, where $\left\vert Y_{2}^{-1},\uparrow
\right\rangle $ is fully occupied while $\left\vert Y_{2}^{1},\uparrow
\right\rangle $ is fully unoccupied. Thus the first order perturbation of
total energy can be written as:

\begin{equation}
E=-\frac{\lambda }{2}\left\vert \cos \theta \right\vert
\end{equation}

The calculated MAE of 140 meV is in good agreement with the expected value
of $\frac{\lambda }{2}$, as the SOC strength $\lambda $ of the 5$d$
electrons is generally regarded as 0.3$\sim $0.5 eV.

\section{Conclusions}

In conclusion, using first-principles and perturbation theory, we present a
comprehensive investigation of the $5d$\ transition-metal oxides Na$_{4}$IrO$%
_{4}$, where Ir occupies the center of square-planar coordination
environment. We discuss its electronic structures, determine its magnetic
ground state configuration, and find a giant MAE for this compound. We
clarify the microscopic mechanism about this novel magnetic properties, and
also suggest possible way to further enhance MAE. We expect that the $5d$
transition metal oxides with low symmetry and long $5d$-$5d$ distance may
exhibit extraordinarily large coercive fields.\ The prediction about giant
MAE deserves experimental tests and may provide a route to nanoscale
magnetic devices.

This work was supported by the Ministry of Science and Technology of China
(Grant number: 2017YFA0303200), NSFC under Grants No. 11525417 and No.
11374137.

\section{Appendix}

\subsection{Symmetry analysis}

Symmetry will add some restrictions on the magnetic model. Considering the
time reversal symmetry and to the quadratic terms of spins, the magnetic
model can be written in the following general form: 
\begin{equation}
H=\frac{1}{2}\sum_{ls,l^{\prime }s^{\prime }}S_{ls}^{\dag }J(ls,l^{\prime
}s^{\prime })S_{l^{\prime }s^{\prime }},  \label{gmm}
\end{equation}%
where $S_{ls}$ represents the magnetic moment located at the magnetic ion
labeled by $s$ in the $l$th unit cell. $J(ls,l^{\prime }s^{\prime })$ is the
exchange interaction between $S_{ls}$ and $S_{l^{\prime }s^{\prime }}$. It
is obviously a $3\times 3$ real matrix, because the magnetic moment is a
three-component vector and we adopt the conventional Cartesian coordinate
system. Translation symmetry will restrict $J(ls,l^{\prime }s^{\prime })$ to
be related to $l^{\prime }-l$, irrespective of the starting unit cell.
Rotation inversion or the combination of two will also give some
restrictions on the exchange matrix. Considering a general space group
element, $\{\alpha |\mathbf{t}\}$, of which $\alpha $ is the point
operation, we denote the representation matrix as $R(\alpha )$ of $\alpha $
in the coordinate system here. Then $J(ls,l^{\prime }s^{\prime })$ should
satisfy that, 
\begin{equation}
R(\alpha )^{\dag }J(mp,m^{\prime }p^{\prime })R(\alpha )=J(ls,l^{\prime
}s^{\prime }),  \label{jres}
\end{equation}%
where $mp$ and $m^{\prime }p^{\prime }$ are related to $ls$ and $l^{\prime
}s^{\prime }$ by the action of $\{\alpha |\mathbf{t}\}$, respectively.%
\newline

We then get ready to turn to the magnetic model for Na$_4$IrO$_4$. Because
there is only one magnetic ion, namely Ir, in one unit cell, we can just
label the magnetic moment by the unit cell label $l$. Utilizing the
translation property, we just need to consider $J(0,l)$, which we denote to
be $J(l)$ or $J(l_1,l_2,l_3)$ hereafter. The onsite exchange $J(0,0,0)$ is
found to own the following form, 
\begin{equation}  \label{000}
J(0,0,0)=\left( 
\begin{array}{ccc}
J(0,0,0)_{11} & 0 & 0 \\ 
0 & J(0,0,0)_{11} & 0 \\ 
0 & 0 & J(0,0,0)_{33}%
\end{array}
\right),
\end{equation}
which represent the single ion anisotropy $\sim S_z^2$.

According to the lattice parameters, we find that $l=(0,0,1)$ and $(0,0,-1)$
are for the nearest-neighborhoods, $l=(\eta_11/2,\eta_21/2,\eta_31/2)$($%
\eta_i=\pm$) for the eight next-nearest-neighborhoods, and $%
l=(\pm1,0,0),(0,\pm1,0)$ for the 3rd-nearest-neighborhoods.

Then $J(0,0,\pm1)$ are found to be in the following form: 
\begin{equation}  \label{001}
J(0,0,\pm1)=\left( 
\begin{array}{ccc}
J(0,0,1)_{11} & 0 & 0 \\ 
0 & J(0,0,1)_{11} & 0 \\ 
0 & 0 & J(0,0,1)_{33}%
\end{array}
\right),
\end{equation}
and note that in the maintext, we relabel $J(0,0,1)_{11}$ and $J(0,0,1)_{22} 
$ to be $J_1^{xx}$ and $J_1^{yy}$, respectively.

For the next-nearest-neighborhoods, inversion symmetry will restrict $%
J(1/2,1/2,1/2)$ to be a symmetric matrix, which would allow finite
non-diagonal elements. However we can ignore these symmetric non-diagonal
elements, because physically they are relatively small \cite{nondiagonal}.
Then the $J(\eta _{1}1/2,\eta _{2}1/2,\eta _{3}1/2)$ are in the following
form, 
\begin{widetext}
 \begin{equation}\label{2221}
    J(\eta_11/2,\eta_11/2,\pm1/2)=\left(
    \begin{array}{ccc}
    J(1/2,1/2,1/2)_{11}& 0& 0\\
    0& J(1/2,1/2,1/2)_{22}& 0\\
    0& 0& J(1/2,1/2,1/2)_{33}\end{array}
    \right),
 \end{equation}
 \end{widetext}%
\begin{widetext}
  \begin{equation}\label{2221}
    J(\eta_11/2,-\eta_11/2,\pm1/2)=\left(
    \begin{array}{ccc}
    J(1/2,1/2,1/2)_{22}& 0& 0\\
    0& J(1/2,1/2,1/2)_{11}& 0\\
    0& 0& J(1/2,1/2,1/2)_{33}\end{array}
    \right),
 \end{equation}
 \end{widetext}and note that in the main text, we relabel $%
J(1/2,1/2,1/2)_{11}$, $J(1/2,1/2,1/2)_{22}$, and $J(1/2,1/2,1/2)_{33}$ to be 
$J_{2}^{xx}$, $J_{2}^{yy}$ and $J_{2}^{zz}$, respectively.

Finally $J(\pm 1,0,0)$ and $J(0,\pm 1,0)$ are found to be in the following
form: 
\begin{equation}
J(\pm 1,0,0)=\left( 
\begin{array}{ccc}
J(1,0,0)_{11} & J(1,0,0)_{12} & 0 \\ 
J(1,0,0)_{12} & J(1,0,0)_{22} & 0 \\ 
0 & 0 & J(1,0,0)_{33}%
\end{array}%
\right) ,  \label{100}
\end{equation}%
\begin{equation}
J(0,\pm 1,0)=\left( 
\begin{array}{ccc}
J(1,0,0)_{22} & -J(1,0,0)_{12} & 0 \\ 
-J(1,0,0)_{12} & J(1,0,0)_{11} & 0 \\ 
0 & 0 & J(1,0,0)_{33}%
\end{array}%
\right) ,  \label{010}
\end{equation}%
and note that in the main text, we relabel $J(1,0,0)_{11}$, $J(1,0,0)_{22}$
and $J(1,0,0)_{33}$ to be $J_{3}^{xx}$, $J_{3}^{yy}$ and $J_{3}^{zz}$,
respectively. The non-diagonal elements $J(1,0,0)_{12}$ are still thought to
be very small and ignored \cite{nondiagonal}.

\subsection{Energy-mapping analysis}

We evaluate spin exchange parameters $J_{1}$-$J_{3}$ by energy-mapping
analysis. Firstly we consider four magnetic configurations as shown in Fig.
4. The total energies of these four spin states can be described in terms of
the spin Hamiltonian:

\begin{equation}
H=\sum_{i<j}J_{ij}\hat{S}_{i}\cdot \hat{S}_{j}
\end{equation}

where $J_{ij}$ (= $J_{1}$, $J_{2}$, $J_{3}$) is the spin exchange parameter
between the spin sites $i$ and $j$. By applying the energy expressions
obtained for a spin dimer with S = 3/2 for Ir$^{4+}$ ions, the total
energies per unit cell for these four spin configurations are expressed as

\begin{align}
E_{FM}& =(2J_{1}+8J_{2}+4J_{3})(\frac{S^{2}}{2}) \\
\quad E_{AFM1}& =(2J_{1}-8J_{2}+4J_{3})(\frac{S^{2}}{2}) \\
\quad E_{AFM2}& =2J_{1}\times (\frac{S^{2}}{2}) \\
\quad E_{AFM3}& =(-\text{ }2J_{1}+4J_{3})(\frac{S^{2}}{2})
\end{align}

The relative energies of the four spin states are obtained from LSDA and
LSDA + $U$ ($U$ = 2 eV) calculations and the values of $J_{1}$ to $J_{3}$
can be evaluated by mapping these energies. The calculated magnetic moments
for the four magnetic configurations have little difference, as shown in
Table. \ref{afm123}. The calculated spin exchange coupling parameters $J_{1}$
to $J_{3}$ are summarized in Table. \ref{Jvalue}. The spin exchanges $J_{1}$%
, $J_{2}$ and $J_{3}$ are all AFM and $J_{2}$ dominates over others in
strength, while $J_{3}$ is almost negligible.

\begin{figure}[tbp]
\centering\includegraphics[width=8cm]{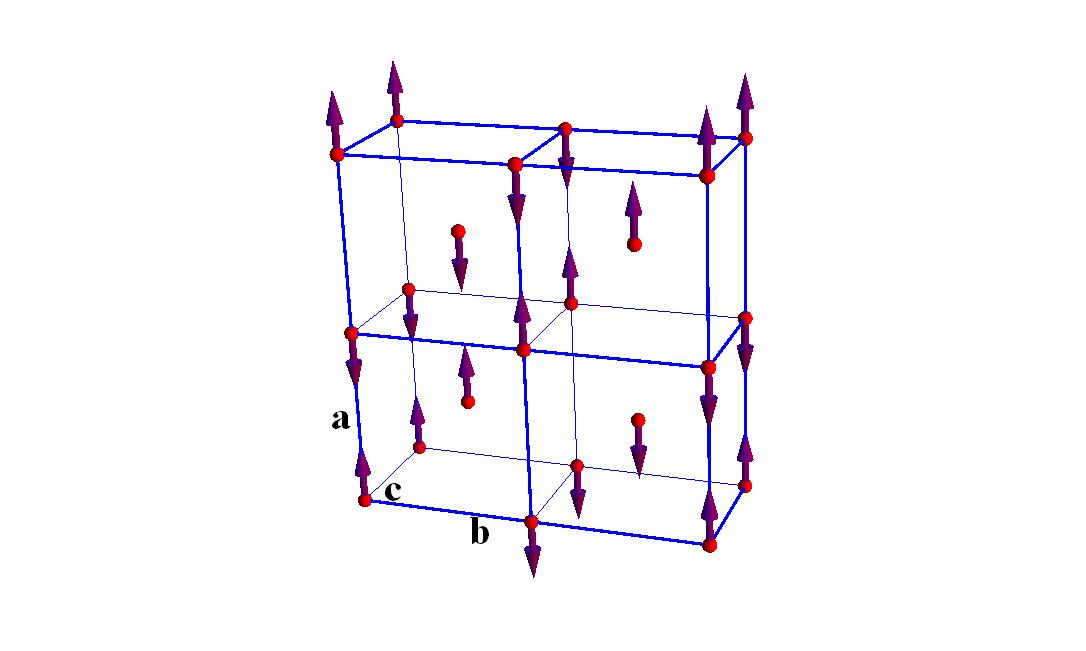}
\caption{The AFM-4 magnetic configuration of Na$_{4}$IrO$_{4}$. For clarity
only Ir atoms are shown.}
\label{app1}
\end{figure}

Considering a generalized symmetry allowed spin model described in Eq. (\ref%
{J}), which includes the anisotropic part of $J$ and single-ion anisotropy,
the values of $J^{xx}$, $J^{yy}$, $J^{zz}$ and $K$ can be determined by
energy-mapping analysis of LSDA + $U$ + SOC calculations with different
magnetization directions. In order to estimate these values, one more
magnetic configuration should be considered, as shown in Fig. \ref{app1}.
The total energies per unit cell with different magnetic configurations are
expressed as

\begin{align}
E_{FM}^{(001)}& =(2J_{1}^{zz}+8J_{2}^{zz}+4J_{3}^{zz})(\frac{S^{2}}{2}%
)-KS^{2} \\
\quad E_{AFM1}^{(001)}& =(2J_{1}^{zz}-8J_{2}^{zz}+4J_{3}^{zz})(\frac{S^{2}}{2%
})-KS^{2} \\
\quad E_{AFM2}^{(001)}& =2J_{1}^{zz}\times (\frac{S^{2}}{2})-KS^{2} \\
\quad E_{AFM3}^{(001)}& =(-\text{ }2J_{1}^{zz}+4J_{3}^{zz})(\frac{S^{2}}{2}%
)-KS^{2} \\
E_{AFM4}^{(001)}& =(2J_{1}^{zz}-4J_{3}^{zz})(\frac{S^{2}}{2})-KS^{2} \\
E_{FM}^{(100)}&
=(2J_{1}^{xx}+4J_{2}^{xx}+4J_{2}^{yy}+2J_{3}^{xx}+2J_{3}^{yy})(\frac{S^{2}}{2%
}) \\
E_{AFM1}^{(100)}&
=(2J_{1}^{xx}-4J_{2}^{xx}-4J_{2}^{yy}+2J_{3}^{xx}+2J_{3}^{yy})(\frac{S^{2}}{2%
}) \\
E_{AFM2}^{(100)}& =(2J_{1}^{xx}+2J_{3}^{xx}-2J_{3}^{yy})(\frac{S^{2}}{2}) \\
E_{AFM3}^{(100)}& =(-2J_{1}^{xx}+2J_{3}^{xx}+2J_{3}^{yy})(\frac{S^{2}}{2}) \\
E_{AFM4}^{(100)}&
=(2J_{1}^{xx}-4J_{2}^{xx}+4J_{2}^{yy}-2J_{3}^{xx}-2J_{3}^{yy})(\frac{S^{2}}{2%
}) \\
E_{FM}^{(010)}&
=(2J_{1}^{xx}+4J_{2}^{xx}+4J_{2}^{yy}+2J_{3}^{xx}+2J_{3}^{yy})(\frac{S^{2}}{2%
}) \\
E_{AFM1}^{(010)}&
=(2J_{1}^{xx}-4J_{2}^{xx}-4J_{2}^{yy}+2J_{3}^{xx}+2J_{3}^{yy})(\frac{S^{2}}{2%
}) \\
E_{AFM2}^{(010)}& =(2J_{1}^{xx}-2J_{3}^{xx}+2J_{3}^{yy})(\frac{S^{2}}{2}) \\
E_{AFM3}^{(010)}& =(-2J_{1}^{xx}+2J_{3}^{xx}+2J_{3}^{yy})(\frac{S^{2}}{2}) \\
E_{AFM4}^{(010)}&
=(2J_{1}^{xx}+4J_{2}^{xx}-4J_{2}^{yy}-2J_{3}^{xx}-2J_{3}^{yy})(\frac{S^{2}}{2%
})
\end{align}

The relative total energies of these spin states are obtained from LSDA + $U$
+ SOC calculations, which are summarized in Table. \ref{app2}. The
calculated magnetic moments for these magnetic configurations have little
difference, as also summarized in Table. \ref{app2}. By energy-mapping
analysis, the calculated anisotropic spin exchange coupling parameters $%
J_{i}^{\alpha \alpha }$ and single-ion anisotropy parameter $K$ are
summarized in Table. \ref{Jvalue}.

\begin{table*}[tbp]
\caption{The calculated total energy (in meV) per unit cell and magnetic
moments (in $\protect\mu $B) for the four magnetic configurations from LSDA
+ SOC + $U$ ($U$ = 2 eV) calculations with (001), (010) and (100) spin
orientations for energy-mapping analysis.}
\label{app2}%
\begin{tabular}{llll|lll|lll|lll|lll}
\hline\hline
& \multicolumn{3}{c}{FM} & \multicolumn{3}{|c}{AFM-1} & \multicolumn{3}{|c}{
AFM-2} & \multicolumn{3}{|c|}{AFM-3} & \multicolumn{3}{|c}{AFM-4} \\ 
& (001) & (100) & (010) & (001) & (100) & (010) & (001) & (100) & (010) & 
(001) & (100) & (010) & (001) & (100) & (010) \\ \hline
E$_{total}$ & 22.3 & 34.4 & 34.4 & 0 & 11.6 & 11.6 & 11.1 & 22.8 & 22.8 & 8.9
& 21.5 & 21.5 & 10.9 & 23.3 & 21.7 \\ 
m$_{Ir}$(spin) & 1.37 & 1.38 & 1.38 & 1.34 & 1.35 & 1.35 & 1.36 & 1.36 & 1.36
& 1.36 & 1.36 & 1.36 & 1.36 & 1.36 & 1.36 \\ 
m$_{Ir}$(orbital) & 0.11 & 0.10 & 0.10 & 0.10 & 0.10 & 0.10 & 0.10 & 0.10 & 
0.10 & 0.10 & 0.10 & 0.10 & 0.10 & 0.10 & 0.10 \\ 
m$_{O}$ & 0.25 & 0.24 & 0.24 & 0.24 & 0.23 & 0.23 & 0.24 & 0.24 & 0.24 & 0.24
& 0.24 & 0.24 & 0.24 & 0.24 & 0.24 \\ \hline\hline
\end{tabular}%
\bigskip
\end{table*}

\subsection{Details of results for Na$_{4}$ReO$_{4}$}

The phonon calculation is performed from the finite displacement method as
implemented in the Vienna ab-initio simulation package (VASP) \cite%
{vasp1,vasp2,vasp3} and the PHONOPY package \cite{vasp4}. After a series of
tests, a $3\times 3\times 2$ supercell is constructed to ensure the force
convergence, and a $2\times 2\times 4$ k-mesh for the Brillouin zone
sampling is used in the phonon calculation. The calculated phonon spectrum
along high-symmetry lines is shown in Fig. \ref{phonon}.

In Na$_{4}$ReO$_{4}$ case, Re$^{4+}$ ion has two less occupied $5d$
electrons than Ir$^{4+}$ ion. We perform first-principles calculations and
find that the crystal field splitting does not change. The calculated
magnetic moment is 0.51 $\mu _{B}$. We perform several calculations for
different magnetic configurations and find that the magnetic ground-state
configuration is the FM state. The calculated spin exchange coupling
parameters $J_{1}$ to $J_{3}$ are -1.84 meV, -0.84 meV and -0.04 meV,
respectively. However, the single-ion anisotropy has a overwhelmingly major
contribution on MAE especially in Na$_{4}$ReO$_{4}$, which has an order of
magnitude larger single-ion anisotropy than Na$_{4}$IrO$_{4}$.

\end{document}